\documentclass{article}
\usepackage{amsmath}

\input{tcilatex}

\begin{document}

\title{Engineering atom-atom thermal entanglement via two-photon process}
\author{Y. Q. Guo, L. Zhou, and H. S. Song \\
\textit{Dalian Univ. of Tech., Dalian, 116024, P. R. China}}
\maketitle

\begin{abstract}
We study the system that two atoms simultaneously interact with a
single-mode thermal field via different couplings and different spontaneous
emission rates when two-photon process is involved. It is found that we
indeed can employ the different couplings to produce the atom-atom thermal
entanglement in two-photon process. The different atomic spontaneous
emission rates are also utilizable in generating thermal entanglement. We
also investigate the effect of the cavity leakage. To the initial atomic
state $|ee\rangle ,$a slight leakage can relieve the restriction of
interaction time and we can obtain a large and steady entanglement.

\ PACS number: 03.67.-a, 03.67.-Hz, 42.50.-p
\end{abstract}

\baselineskip=19pt

\section{Introduction}

Entanglement plays an important role in respect that it is a valuable
resource in quantum information processing such as quantum teleportation\cite%
{1}, quantum computation\cite{2} and quantum cryptography\cite{3}, \textit{%
etc}. Several schemes have been proposed to prepare purified and distilled
entangled state both theoretically and experimentally\cite{4}. Although the
interaction between a quantum system and its surroundings can result in
inevitable decoherence of the quantum system, people have recognized that we
can employ the interaction to generate entanglement\cite{5}.

The two-atom entangled states are widely studied in cavity QED\cite%
{6,7,8,9,10}. In cavity QED, the dissipation in the model of atoms
interacting with magnetic field generally includes two aspects: the cavity
leakage through which the intra-cavity magnetic field can exchange
information with its environmental noise, the atomic spontaneous emission
that is induced by vacuum fluctuation effect. In the sense of using the
impact of environmental noise, the noise-assisted entanglement schemes have
been put forward by many authors\cite{9,10,11,12,13,14,15}. Plenio and
co-work have developed schemes that involves continuous monitoring of
photons leaking out of the cavity to entangle atoms one of which is
initially exicted\cite{9}. In Ref. [10], the author studied the interaction
of a thermal field with a two-qubit system that initially prepared in
separable states. They demonstrated that entanglement of atom-atom can arise
depending on initial preparation of the atoms. Also in Ref. [11], the
entanglemet of atom-atom can be generated through interaction of atoms with
cavity mode coupled to a white noise. Their entanglement can be maximized
for intermediate value of noise intensity and initial value of spontaneous
rate. In these studies, the couplings of atoms to field are confined to be
equal. In fact, the coupling rate $g$ between atomic internal levels and the
cavity mode depends on the atom's position $\mathbf{r}(x,y,z)$ \cite{16}.
The atoms can not be localized precisely even by employing cooling
technology and trapping potential schemes. So, it is practically necessary
to address the question: how will the entanglement be when two atoms
differently couple to a single model field? In Ref.\cite{14}, our gruop had
shown that different couplings can really assist the induce of entanglement
in one-photon process.

On the other hand, the atomic spontaneous emission rate is also related to
atoms's position\cite{16}. In real experimental scenario, the atoms's
position $\mathbf{r}(x,y,z)$ not only dominates the atom's coupling strength
to the field, but also determines the amount of atomic spontaneous emission
rate. It has already been reported that the resonant cavity which was made
of two spherical niobioum mirrors can enhance or suppress single atomic
spontaneous emission by adjusting atom position $Z$ (the distance from
median plane of cavity)\cite{17,18}. But theoretically, atomic spontaneous
emissions have been assumed to be equal or even been ignored, and the
spontaneous emission has been disliked because of its impact on the
entanglement\cite{12}. Up to now we have not found the study that two atoms
spontaneous emission rates are not the same. Addition to that, the
two-photon process is a kind of important one which may show different
properties from the case of one photon in quantum information processing,
for example, it has been found that the atom-atom entanglement induced by
thermal field in two-photon process is larger than that in one-photon process%
\cite{19}.

In this paper, considering the two-photon process, we\ aim to study the two
atoms simultaneously interacting with a single-mode cavity field with
different couplings and different spontaneous emission rates. We find that
in two-photon process we indeed can employ the different couplings to
produce the the atom-atom thermal entanglement. If the atoms spontaneously
emit inevitably, the different spontaneous emission rates is utilizable in
generating thermal entanglement. We also investigate the effect of the
cavity leakage. To the initial atomic state $|gg\rangle $, the cavity
dissipation should be supressed as possible as we can, but to the initial
atomic state $|ee\rangle ,$we can keep a slight leakage to relieve the
restriction of interaction time so that we can obtain a large and steady
entanglement.

\section{The interaction of two-atom system and the Master equation}

The two two-level identical atoms (atom $a$ and atom $b$) are supposed to
interact with a single mode cavity field which is in a thermal equilibrium
with its environment characterized in terms of an mean photon number $N=(e^{-%
\frac{\hbar \omega }{\kappa _{B}T}}-1)^{-1}$, and $T$ is the environmental
temperature. We assume the excited atom can transit from its upper state to
its lower state and emit two photons. So that, the atomic transition
frequency $\omega _{0}$ doubles the field frequency $\omega $. The \textit{%
Hamiltonian }under the rotating wave approximation is

\begin{equation}
H=\omega _0\sigma _a^z+\omega _0\sigma _b^z+\omega
a^{+}a+\sum\limits_{i=a,b}g_i(a^2\sigma _i^{+}+a^{+2}\sigma _i^{-})\text{.}
\end{equation}
Where $a$ and $a^{+}$ represents annihilation and creation operator of
cavity mode respectively. The operators $\sigma _i^{-}$ and $\sigma _i^{+}$
denote atomic transition operators of atom $i$. The coupling constant for
two-photon transition between atom $i$ and the cavity mode is $g_i$. In the
interaction picture, the \textit{Hamiltonian }is

\ \ \ \ \ 
\begin{equation}
H_I=g_aa^2\sigma _a^{+}+g_ba^{+2}\sigma _b^{-}\text{.}
\end{equation}
For the sake of the two couplings' diversity, the following transformation
is preferred

\begin{equation}
g=\frac{g_{a}+g_{b}}{2}\text{, }r=\frac{g_{a}-g_{b}}{g_{a}+g_{b}}\text{,}
\end{equation}
where the $r$ is in the range of $0$ and $1$.

For generality, we assume the intra-cavity system can exchange information
with thermal environment due to cavity dissipation and atomic spontaneous
emission.The time evolution of the global system (atoms+cavity mode) is
governed by the master equation

\begin{equation}
\dot{\rho}=-i[H,\rho ]+L(\rho )\text{.}
\end{equation}%
The \textit{Liouvillean} that describes the atomic spontaneous emission and
the interaction of the cavity mode with the thermal environment in a leaky
cavity is written as\cite{20}

\begin{eqnarray}
L(\rho ) &=&-\sum_{i=a,b}[(\bar{n}+1)\Gamma _{i}(\sigma _{i}^{+}\sigma
_{i}^{-}\rho +\rho \sigma _{i}^{+}\sigma _{i}^{-}-2\sigma _{i}^{-}\rho
\sigma _{i}^{+})  \notag \\
&&n\Gamma _{i}(\bar{\sigma}_{i}^{-}\sigma _{i}^{+}\rho +\rho \sigma
_{i}^{-}\sigma _{i}^{+}-2\sigma _{i}^{+}\rho \sigma _{i}^{-})] \\
&&-\kappa (\bar{n}+1)(a^{+}a\rho +\rho a^{+}a-2a\rho a^{+})  \notag \\
&&-\kappa \bar{n}(aa^{+}\rho +\rho aa^{+}-2a^{+}\rho a)\text{,}  \notag
\end{eqnarray}
where the terms including $\kappa $ in $L(\rho )$ are interpreted as the
coupling strength of cavity mode to the external thermal field, $\Gamma _{i}$
is the spontaneous emission rate of atomic $i(i=a,b)$. Since $\Gamma _{a}$
can be different from $\Gamma _{b}$, we adopt the transformation similar to
Eq. 3 
\begin{equation}
\Gamma =\frac{\Gamma _{a}+\Gamma _{b}}{2}\text{, }\gamma =\frac{\Gamma
_{a}-\Gamma _{b}}{\Gamma _{a}+\Gamma _{b}}\text{.}
\end{equation}
\ \ 

\ The Wootters concurrence that has been proved effective in presenting the
entanglement degree of two qubits is written as\cite{21}

\begin{equation}
C=\max \{0,\lambda _{1}-\lambda _{2}-\lambda _{3}-\lambda _{4}\}
\end{equation}%
where the $\lambda _{i}$ are non-negative real square roots of the
eigenvalues of the \textit{Hermitian} matrix $\sqrt{\rho }\tilde{\rho}\sqrt{%
\rho }$ in decreasing order with $\tilde{\rho}=(\sigma _{y}\otimes \sigma
_{y})\rho ^{\ast }(\sigma _{y}\otimes \sigma _{y})$. No matter what $\rho $
stands for a pure or a mixed entangled state, Wootters concurrence is
available. The amount of entanglement measured by concurrence on the basis
of different initial atomic states will be numerical calculated in next two
sections.

\section{Atom-atom thermal entanglement under different couplings and
different spontaneous emission rates}

We assume that the single mode cavity field is initially in a thermal field
state. Due to the cavity leakage when the cavity is in a thermal equilibrium
with its environment, the cavity field is in a mixture of Fock states. So,
the cavity field initially takes the form

\ \ \ \ \ \ \ \ \ \ \ 
\begin{equation}
\rho _{f}(0)=\sum_{n}\left| n\right\rangle \left\langle n\right| \frac{N^{n}%
}{(1+N)^{n+1}}\text{.}
\end{equation}

Firstly, we study the effect of relative coupling difference $r$ on the
two-atom entanglement when two-photon process is involved. The chosen
parameters are $g=1$, $N=1.5$, $\Gamma _{i}=0$ and $\kappa =0$. We have cut
off the intra-cavity photon number at a value of $5$ which is precise enough
in respect that $\frac{N^{n}}{(1+N)^{n+1}}$ is a decreasing function of
photon number. Fig.1a shows the entanglement as a function of relative
coupling difference $r$ and time $t$ in the case of the two atoms are
initially in $|ee\rangle $, and Fig.1b is the same as Fig. 1a except that
the two-atom are initially in $|gg\rangle $. In Ref.[14], the authors could
not find the entanglement induced by thermal field in two-photon process
when the initial atomic state is $|ee\rangle $ if two couplings are equal.
Fig. 1a also shows there is no entanglement when $r=0$. But if $r\neq 0,$in
some region one can find entanglement. So, in two photon process the
different couplings can also benefit to produce entanglement. Comparing
Fig.1a with Fig.2\ of Ref.[14], we find that the entanglement in one-photon
process (Fig.2 of Ref.[14]) exists in some discontinuous small areas in
terms of $r$ and $t$, i.e., for different $r$ entanglement may appear in
different interval of time, however, Fig. 1a shows that the entanglement
appears in some continuous regions, that is to say, in the relative slowly
varying region of time the entanglement keeps its value even the relative
large change of $r$. This property will be more obvious when the two-atom
are initially in $|gg\rangle $, which is shown in Fig. 1b. The behavior that
the entanglement varies with $r$ and $t$ is very interesting, and the
entanglement can exist in almost the same interval of time for different $r$%
. For example, in the region $0.8<t<1.4$, the entanglement increases to a
maximum slightly with the increasing of $r$ from $0$ to $0.8$, then it
decreases to zero. In other words, if we control the interaction time in the
interval $0.8<t<1.4$, we need not have to care much about whether the two
atoms are in the same position or not. So, in two-photon process it is
experimentally not necessary to control the position precisely, especially
when the initial atomic state is $|gg\rangle $.

Then, we consider the effect of different spontaneous emission rates of
two-atom on the amount of entanglement. We show the typical result of
atom-atom entanglement as a function of the difference between two emissions
and the noise intensity (mean photon number of thermal environment) in Fig.
2a and Fig. 2b corresponding to atomic initial states $\left|
ee\right\rangle $ and $\left| gg\right\rangle $ respectively. The chosen
parameters are $\kappa =0$, $g=1$, $r=0.3$, $\Gamma =0.2$ and $t=1$ in Fig.
2a and $\kappa =0$, $g=1$, $r=0.3$, $\Gamma =0.02$ and $t=1$ in Fig. 2b.
From Fig. 2, we see that the amount of entanglement when difference of two
emissions equals to zero, i.e. $\Gamma _{a}=\Gamma _{b}$, is not the best
case of atom-atom entanglement. The maximum value of entanglement is
monotonously increased by increasing the relative difference of two
spontaneous emissions $\gamma $. For example, the entanglement when $\gamma
=1$ in Fig. 2a is about $1.5$ times of that when $\gamma =0$, and in Fig.
2b, the entanglement when $\gamma =1$ is even enhanced to be about $6$ times
of that when $\gamma =0$. And Fig.2a shows that the entanglement decreases
monotonously with the increasing of mean photon number which is also
observed in Ref.[14]. One can also observe that the entanglement can be
increased by increasing mean photon number in some region in Fig. 2b. This
is because that the two atoms initially in $\left| gg\right\rangle $ can not
be entangled when they interact with vacuum field state. With the increasing
of mean photon number in some extent, the entanglement is gradually
increased to a maximum. One can observe that the amount of entanglement with
spontaneous emission is quite different from that in Fig. 1(without
spontaneous emission). When there is atomic spontaneous emission, even this
emission is very weak, the amount of entanglement will be much weakened. As
mentioned above, in any experimental scenario, the atomic spontaneous
emissions can hardly be all kept as zero. Therefore, any entanglement that
has been realized experimentally is in fact smaller than theoretical result
of ideal model. To investigate the influence of atomic spontaneous emission
on the atom-atom entanglement, the authors in Ref. [12] assume two atoms
have a same spontaneous emission $\Gamma $ in a vacuum cavity. Their results
show that the amount of entanglement is a monotone decreasing function of $%
\Gamma $. While, if there is inevitable spontaneous emission in experiment,
the difference of spontaneous emission rates can also assist atom-atom
entanglement.

\section{The effects of dissipation on the atom-atom thermal entanglement}

We now turn to consider the situation when cavity keeps on leaking
throughout the whole evolution. Fig.3 shows the atom-atom entanglement
changing with cavity dissipation $\kappa $ and time $t$. Fig. 3a and Fig. 3b
are corresponding to the initially atomic state$|gg\rangle $ and$\left|
eg\right\rangle $ respectively. The chosen parameters in both cases are $%
N=1.5$, $\Gamma _{a}=\Gamma _{b}=0$ and $r=0.1$. When the two-atom are
initially in $|gg\rangle $, the amount of entanglement is a monotone
decreasing function of cavity decay. With the cavity dissipation increasing,
the entanglement decreases rapidly. It denotes that we should depress the
cavity dissipation as possible as we can if the initial atomic state is $%
|gg\rangle .$ However, when the initial atomic state is $\left|
eg\right\rangle $, a slight increasing of $\kappa $ makes the period of
entanglement disappears and futhermore benefits to generate relative steady
and strong entanglement. Although the entanglement may decrease slightly as
time evolution, we still can employ the non-period to relieve the
restriction of interaction time. In experiment, precisely controlling
interaction time is still very difficult. While, it will be not necessary to
precisely control the interaction time by employing the slight cavity
dissipation. Thus, the dissipation of the cavity is not always bad to the
atom-atom thermal entanglement. In some cases such as the initial atomic
state $\left| eg\right\rangle $, the cavity dissipation is utilizable.

\section{Conclusion}

To sum up, when two-photon process is involved, we study two atoms
simultaneously interact with the thermal field under different couplings and
different spontaneous emission rates. To different initial atomic state, the
different couplings assist to produce the the atom-atom thermal entanglement
in two-photon process. This entanglement is more regular than that of one
photon process in sense that in some time intervals the entanglement can
survive when difference of two couplings varies in a large range. If the
atoms spontaneously emit inevitably, the different spontaneous emission
rates is utilizable in generating thermal entanglement. The different
spontaneous emission rates can be realized experimentally by localize
different atoms in different places in a same F-P cavity. We also
investigate the effect of the cavity leakage. To the initial atomic state $%
|gg\rangle $, the cavity dissipation should be supressed as possible as we
can, but to the initial atomic state $|ee\rangle $, we can employ a slight
cavity leakage to relieve the restriction of interaction time so that we can
obtain a large and steady entanglement.

{\huge Figure Captions:}

Fig. 1a: Two-atom entanglement versus difference of two couplings $r$ and
time $t$ for atomic initial state $\left| ee\right\rangle $, $N=1.5$, $%
\kappa =0$ and $\Gamma _{a}=\Gamma _{b}=0$.

Fig. 1b: Descriptions are same as in Fig. 1a but for atomic initial state $%
\left| gg\right\rangle $.

Fig. 2a: Two-atom entanglement versus difference of two spontaneous emission
rates $\gamma $ and mean photon number $N$ for atomic initial state $\left|
ee\right\rangle $, $g=1$, $r=0.3$, $\kappa =0$, $t=1$ and $\Gamma =0.2$.

Fig. 2b: Descriptions are same as in Fig. 3a but for $\Gamma =0.02$, $r=0.1$
and atomic initial state $\left| gg\right\rangle $.

Fig. 3a: Two-atom entanglment versus cavity decay $\kappa $ and time $t$ for
atomic initial state $\left| gg\right\rangle $, $g=1$, $r=0.1$, $N=1.5$, $%
\Gamma _{a}=\Gamma _{b}=0$.

Fig. 3b: Descriptions are same as in Fig. 3a but for atomic initial state $%
\left| eg\right\rangle $.

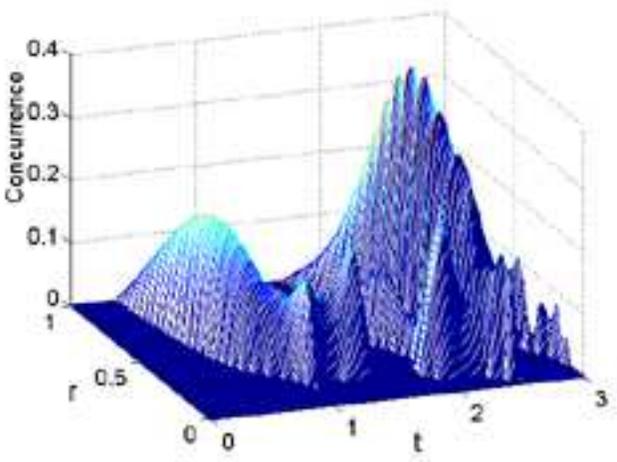
\end{document}